\pdfoutput=1
\documentclass[11pt]{article}
\usepackage{graphicx}
\usepackage{amsmath}
\usepackage{amssymb}
\usepackage{amsxtra}

\title{Small Representation Principle}
\author{H.B.~Nielsen\thanks{e-mail: hbech@nbi.dk, hbechnbi@gmail.com}\\
The Niels Bohr Institute, Copenhagen,
Denmark} 
\begin{document}

\maketitle

\begin{abstract}
  In a previous article \cite{seeking} Don Bennett and I 
looked for, found and proposed a game in which the Standard Model
Gauge {\em Group} $S(U(2) \times U(3))$ gets singled out as the
``winner''.  This ``game'' means that the by Nature chosen gauge group
should be just that one, which has the maximal value for a quantity,
which is a modification of the ratio of the quadratic Casimir for the
adjoint representation and that for a ``smallest'' faithful
representation.  In a recent article \cite{dimension} I proposed to
extend this ``game'' to construct a corresponding game between
different potential dimensions for space-time.  The idea is to
formulate, how the same competition as the one between the potential
gauge groups would run out, if restricted to the potential Lorentz or
Poincare groups achievable for different dimensions of space-time
$d$. The remarkable point is, that it is the experimental space-time
{\em dimension 4, which wins}.

Our ``goal quantity'' to be maximized has roughly the 
favouring meaning that the Lie-group in question can have the
``smallest'' possible faithful representations. This idea then
suggests that the representations of the Standard Model group to be
found on the (Weyl)Fermions and the Higgs Boson should be in the
detailed way measured by our ``goal quantity'' be the smallest
possible. The Higgs in the Standard Model belongs remarkably enough
just to the in such a way ``smallest'' representation. For the chiral
Fermions there are needed restriction so as to avoid anomalies for the
gauge symmetries, and in an earlier work\cite{CostaRica,parity} we
have already suggested that the Standard Model Fermion representations
could be considered being the smallest possible. We hope in the future
to show that also taking smallness in the specific sense suggested
here would lead to the correct Standard Model representation system.

So with the suggestion here the whole Standard Model is specified by
requiring SMALLEST REPRESENTATIONS! Speculatively we even argue
that our principle found suggests the group of gauge transformations
and some manifold(suggestive of say general relativity).
 \end{abstract}

\section{Introduction}
In two earlier 
articles\cite{seeking,dimension} Don 
Bennett and 
I proposed a quantity depending on a 
group - thought of as the gauge 
{\em group}(group in the 
sense of O'Raifeartaigh 
\cite{ORaifeartaigh} 
- which were found 
to take its largest value on just the 
Standard Model gauge {\em group} 
$S(U(2)\times U(3))$. My article 
\cite{dimension} were to tell that 
the same quantity applied to the 
Lorentz or by some crude technology to
essentially the Poincare group selected 
as the number of dimensions winning the 
highest quantity just the experimental 
number of dimensions 4 for space time.
(The prediction of the d=4 dimensions
from various reasons have been considered
in e.g. \cite{Tegmark,Ehrenfest,Kane,Norma}. N.Brene and I have earlier 
proposed another quantity to be extremized
to select the Standard Model group, namely
that it is the most ``skew''\cite{Brene}
(i.e. it has the smallest number of 
automorphisms, appropriately counted).
But in this article we shall discuss 
a ``goal quantity''that  being maximized
as it shall, rather may mean 
crudely that the group can have as 
small representations as possible.

To define this wonderful group dependent
quantity, which can in this way select as 
the highest scoring group the by Nature
chosen Standard Model {\em group}, and 
the by 
Nature chosen space time dimension 4, let 
us think of a general Lie group written 
by means of a cross product of a series 
of simple Lie groups $H_i$ (take the 
$H_i$'s to be  the covering 
groups at first) and a series of real 
number ${\bf R}$ factors in this cross
product
\begin{equation}
G_{cover} =\left ({\hbox{\huge \bf 
$\times$}}_i^I H_i\right )
\times  {\bf R}^{J}.
\end{equation}  
Here the I is the number of,  
different or identical, as
it may be, $H_i$-groups, which  are 
supposed to 
be simple Lie groups, while ${\bf R}$
denotes the Abelian group of real 
numbers under addition.The number of 
Abelian dimensions in the Lie algebra 
is called  $J$. A very general 
group is obtained by dividing an invariant
discrete subgroup $D$ of the center out 
of this group
$G_{cover}$. Denoting this general 
 - though assumed connected -
group
as  $G$ we can indeed write it as
\begin{equation}
G = G_{cover}/D. 
\end{equation}
Of course $G_{cover}$ is the covering group
of $G$ and the groups $H_i$ (i = 1,2, ...
I-1,I) are its invariant simple Lie groups.

The main ingredient in defining our goal 
quantity is the ratio of the quadratic 
Casimirs\cite{Casimir}  $C_A/C_F$ of the 
quadratic 
Casimir $C_A$ for the adjoint 
representation divided by the 
quadratic Casimir $C_F$ a representation 
chosen, so as to make the quadratic 
Casimir
$C_F$ 
of $F$ so small as possible though still 
requiring the representation $F$ to be 
faithful or basically to be non-trivial.
Here I now ought to remind the reader 
of the concept of a quadratic Casimir 
operator: 

The easiest may be to remember the concept
of quadratic Casimir first for the 
most well known example of a 
nonabelian Lie group, namely the group 
of rotations in 3 dimensions $SO(3)$ 
(when you do not include reflection in a
point but only true rotations) for which 
the covering group is $SU(2)$. In this 
case the quadratic Casimir operator 
is the well known square of the angular
momentum operator 
\begin{equation}
\vec{J}^2 = J_x^2 + J_y^2 + J_z^2
\end{equation}

Now our goal quantity, which so nicely 
points to both the Standard Model group 
and the dimension of space time,  
is given as the $d_G$'th root of 
the product with one factor from each 
invariant simple group $H_i$,
namely $(C_A/C_F)^{d_i}$ ($C_F/C_A$ is 
related to 
the Dynkin-index \cite{Dynkinindex}) and 
some factors 
$e_A^2/e_F^2$ for each of the $J$ 
Abelian factors.  (Here the dimension 
of the simple groups $H_i$ are denoted 
$d_i$, while the dimension of the 
total group $G$ or of $G_{cover}$ is 
denoted $d_G$.) Our goal quantity 
in fact  becomes
\begin{eqnarray}
&&\hbox{``goal quantity''}\\
&=& \left (\prod_{\hbox{simple groups $i$}}
(\frac
{C_A}{C_F})_i^{d_i} * \prod_{\hbox{Abelian 
factors $j$}}^J(\frac{e_A^2}{e_F^2})_j 
\right )^{1/d_G}.\label{goal}  
\end{eqnarray}
To fully explain this expression I need 
to explain what means the ``charges'' 
$e_F$ for the ``small'' representation
(essentially $F$) and $e_A$ for the 
analogon
\footnote { We might define an 
analogon of the adjoint representation for 
also a set of $J$ properly chosen 
${\bf R}$-factors of $G$ by assigning 
the notion of ``analogon to the adjoint
representation'' to that representation 
of one of the ${\bf R}$-factors, which 
has the smallest charge, $e_A$ called, 
allowed
for a representation of ${\bf R}/({\bf R}
\cap D)$, where ${\bf R}$ 
stands for the ${\bf R}$-factor 
considered, and ${\bf R}\cap D$ for
the intersection of the to be  divided
out discrete group $D$ with this Abelian 
factor {\bf R}.}   
    to the adjoint representation: Of 
course the 
reader  should have in mind that
the Abelian groups, the ${\bf R}$ 
subgroups, have of course no adjoint 
representations in as far as the basis 
in the Lie algebra of an Abelian group 
is only transformed trivially. In stead 
of defining these ``charges'' -- as we 
shall  do below -- by first defining a 
replacement for the adjoint we shall 
define these factors  $\prod_{
\hbox{Abelian 
factors $j$}}^J(\frac{e_A^2}{e_F^2})_j$
from the Abelian factors in the Lie 
algebra by means of the system of 
{\em allowed and not allowed 
representations} 
of the {\em group} $G =G_{cover}/D =
\left(\left ({\hbox{\huge 
\bf $\times$}}_i^I H_i\right )
\times  {\bf R}^{J} \right )/D$. Each 
irreducible representation of this 
$G$ is characterized in addition to 
its representations under the simple
Lie groups $H_i$ also by a ``vector''
of ``charges'' representing the phase 
factors $\exp(i\delta_1e_{r1}
+i\delta_2e_{r2}+ ... + i\delta_Je_{rJ})$, 
which multiply the 
representation vector under an 
element $(\delta_1, \delta_2, ...
,\delta_J)\in {\bf R}^J$, i.e. in the 
Abelian factor of $G$. 
The easiest may be to say that we consider
the whole lattice system of allowed 
``vectors'' $\{ (e_{r1} , e_{r2}, ...,
e_{rJ})|r \hbox{allowed by G}\}$ of sets 
``charges'' allowed by the {\em group}
$G$, and then compare with corresponding
set in which we {\em only consider those 
representations $r$, which represent
the simple non-abelian groups only 
trivially}:
\begin{eqnarray}
 \{ (e_{r1} , e_{r2}, ...,e_{rJ})|r &&\hbox{allowed by $G$,}\nonumber\\
 &&\hbox{ and with the 
representation of the $H_i$'s being 
only trivial}\}.\nonumber
\end{eqnarray}
In this comparison
you ask for a going to an infinitely big 
region 
in the $J$-dimensional lattice after 
the ratio 
of the number of ``charge vectors''
in the first lattice 
$$\{ (e_{r1} , e_{r2}, 
...,e_{rJ})|r \hbox{allowed by G}\}$$ 
relative
to that in the second 
$$\{ (e_{r1} , e_{r2}, ...,e_{rJ})|r \hbox{ $G$-allowed with 
the  $H_i$'s represented trivially}\}.$$ 
Then the whole 
factor under the $d_G$'th root sign
is the product of the factor comming 
from the semisimple part of the 
group $G$ 
\begin{equation}
\hbox{``Semisimple factor''}
=\prod_{\hbox{simple groups $i$}}
(\frac
{C_A}{C_F})_i^{d_i}\label{semisimplef}
\end{equation}       
and the ratio of the number of 
charge combinations  at all allowed by the 
{\em group} $G$ to the number of 
charge combinations, when the semisimple 
groups are restricted to be represented 
trivially - in the representation 
of the whole $G$ representing the 
Abelian part by the charge combination 
in question :
\begin{eqnarray}
\hbox{``Abelian factor''}=\nonumber\\
\left (\frac{\#  \{ (e_{r1} , e_{r2}, ...,
e_{rJ})|r \hbox{ allowed by G}\}}{\# \{ (e_{r1} , e_{r2}, ...,e_{rJ})|r \hbox{ $G$-allowed
{\em  with 
the  $H_i$'s represented
 trivially}}\}}\right )^2 .
\end{eqnarray} 
Here $\#$ stands for the number of 
elements in the following set, i.e.
the cardinal number; but it must be 
admitted that the numbers of 
these charge combinations are infinite,
and that to make the finite result,
which 
we shall use, we have to take a cut off 
and take the limit of the ratio for 
that cut off going to be a bigger and 
bigger sphere finally covering 
the whole $J$-dimensional space 
with the charge combinations embedded.
So strictly speaking we define rather
\begin{eqnarray}
&&\hbox{``Abelian factor''}=\nonumber\\
&&\left (\lim_{S \rightarrow \infty}
\frac{\#  \{ (e_{r1} , e_{r2}, ...,
e_{rJ})|r \hbox{ allowed by G}
\}_{\hbox{cut off by S}}}{\# \{ (e_{r1} , e_{r2}, ...,e_{rJ})|r \hbox{ $G$-allowed
{\em  with 
the  $H_i$'s represented
 trivially}}\}_{\hbox{cut off by S}}}
\right )^2. \nonumber
\label{abelianf}
\end{eqnarray} 
where S is some large ``sphere say''
in the $J$-dimensional space of charge 
combinations. The symbol $S \rightarrow
\infty$ shall be understood to mean 
that the region S is taken to be larger 
and  larger in all directions so as to in 
the limit cover the whole space.

Then our goal quantity to be maximized
so as to select the gauge group supposed 
to be chosen by nature can be written
\begin{equation}
\hbox{``goal quantity''}
= \left (\hbox{``Semisimple factor''}
*\hbox{``Abelian factor''} \right
)^{1/d_G}.
\end{equation}

Really it is nice to express the 
quantity $\hbox{``Abelian factor''}$
by means of the representations 
allowed by the group, because after all
the phenomenological determination 
of the {\em Lie-group} rather than 
only the Lie algebra\cite{ORaifeartaigh}  
is based on  
such a system of allowed representations.

\subsection{Motivation}
Before illustrating the calculation of 
our ``goal quantity'' with Standard Model
as the example, let me stress the 
motivation or interest in looking for 
such a function defined on gauge groups 
or more abstractly somehow on theories 
and can be used to single out the by 
Nature chosen model. A major reason 
making such a singling out especially 
called for is that the Standard Model 
and e.g. its group is not in an obvious 
way anything special! It is a combination 
of several subgroups like $SU(2)$, 
$SU(3)$, and $U(1)$ of groups that cannot
all be the obvious {\em one}, since we 
already use 3. There exist both several 
groups with lower rank, say than the 
4 of the Standard Model group, and of 
cause 
infinitely many with higher rank. That it 
truly has been felt, not only by us, but 
by 
many physicists that the Standard Model 
is {\em a priori} not anything obviously 
special
- except for the fact, that it is the 
model that agrees with experiment -
can be seen from the great interest in 
- and even belief in - grand unification
theories\cite{GUT} seeking to find e.g. 
an extended 
gauge group, of which the Standard Model 
gauge group is then only the small part,
which survived some series of (spontaneous)
break downs of part of the larger group.
Let me 
put some of the predictions 
of the  typical grand unification model
as $SU(5)$ in the perspective: When 
they are concerned with representations 
possible for say the $SU(5)$, there are 
restrictions for what they can be for 
the Standard Model ``$SU(2) \times SU(3)
\times U(1)$''- and they agree with 
experiment -;but  then these restrictions 
are truly a 
consequence of that {\em the 
sub{\bf group} 
of $SU(5)$ having the Lie-algebra (of) 
$SU(2)\times SU(3) \times U(1)$ is 
precisely the {\bf group} $S(U(2)\times 
U(3))$}. Indeed the condition on the 
possible representations, when there is 
an $SU(5)$ GUT theory beyond the 
Standard Model, is the same condition
(\ref{restriction}) as comes from  
$S(U(2)\times U(3))$. 
 There is of course more 
information in specifying the group than 
only the Lie-algebra; but that of course 
only implies that an a priori not special
group is even less special than an 
algebra, because there are even more 
{\em groups} among 
which to choose than there are {\em 
algebras}.
(Of course there are truly infinitely 
many both groups and algebras, but for a 
given range of ranks, say, there are more 
Lie-{\em groups} than Lie-{\em algebras}).

Another hope of explaining, why the 
Standard Model including its gauge group
is chosen by Nature,
is the superstring theories, which predict
 at the fundamental or string level the 
gauge groups $E_8\times E_8$ or $SO(32)$.
But from the point of view of our ``goal
quantity'' - as can be seen below form 
our tables - especially $E_8$ and 
consequently also  $E_8\times E_8$ 
(since our ``goal quantity'' has the 
property of being the same for a 
group $G$ and its cross products with 
itself any number of times) is the worst 
group from the point of view of our 
``goal quantity'': In fact the nature 
of our ``goal quantity'' construction 
is so, that  we always must have
\begin{equation}
\hbox{``goal quantity''} \ge 1.
\end{equation}
But $E_8$ according to the table 
below gives just this 1 for its 
``goal quantity'' 
$$\hbox{``goal quantity''}_{E_8\times E_8}
= \hbox{``goal quantity''}_{E_8} =1$$
actually because $E_8$ has no smaller 
representation than its adjoint 
representation.

The connection to my personal pet-theory
(or dream, or program) of Random 
Dynamics \cite{RD,RD2,RDrev,Foerster,Astri,RDDon1} is that a priori the 
present work is ideally phenomenologically
- as to be explained in subsection 
\ref{phen}-,i. e. the spirit is to ask nature and just seek to find what is characteristic for the Standard Model group 
without theoretical guesses behind a 
priori.  However, it(= our phenomenological result) leads to the 
suggestion
that the (gauge)group that wins - gets 
highest ``goal quantity'' - is the one 
that most likely would become  
approximately
a good symmetry by accident. This would 
then mean, that in a random model, as is
the picture in Random Dynamics, the group,
that is selected by our game, is just the 
one most likely to be realized as an 
approximately good symmetry by accident.
So indeed Random Dynamics could be 
a background theory for the present work.
So in this sense random dynamics ends 
up being favoured by the present article,
although we in principle started out 
purely phenomenologically. 
(It must be admitted though, that 
historically the idea appeared as 
an {\em extract} from a long Random 
Dynamics 
inspired calculation - which has so 
far not been published - by Don 
Bennett and myself.) Having approximately
gauge symmetries, there is according to 
some earlier works of ours and others 
\cite{Foerster,Damgaard,Lehto,Lehto2}
the possibility that the gauge symmetry 
may become exact by quantum fluctuations;
really one first writes it formally as if
the remaining small breaking were a 
Higgsing, 
and then argue that quantum fluctuations
wash away this ``Higgs''effect.
\subsection{Plan of Article}
 In the next section \ref{example}
we shall with the Standard Model group
as an example tell how to calculate 
the goal quantity, and we deliver in this 
section \ref{example} also some {\em 
tables} to use for such 
computations. Then in section \ref{dim} 
we discuss the attempt to also postdict 
the dimension of space time; for that 
several slight modifications are used 
to in an  approximate sense construct a
goal quantity like quantity for even the
Poincare group in an arbitrary dimension d
for space time.
Successively in section \ref{Higgs} we 
consider, how we can extend our ideas
to measure the {\em size of a 
representation} 
of the Standard Model group, and then 
the wonderful result is that the 
representation, under which the 
{\em Higgs} 
fields transform, remarkably enough 
turns out to be just the {\em smallest 
(non-
trivial) representation}! 

The following sections are about work 
still under 
development, and in section 
\ref{fermionrep} we review an old 
work making more precise, what is 
already rather intuitively obvious:
That the fermion representations in the 
Standard Model are rather ``small''
and that that together with anomaly 
conditions settles what they can be 
{\em assuming mass protected fermions 
only}. In the next 
section \ref{full} we point to a way 
of changing  the point of view so as to 
say, that, what we predict, is rather than
the gauge group the group of gauge 
transformations. This may be the beginning
to predict also a manifold structure for 
the whole gauge theory. Are we on the 
way to general relativity? We conclude 
and resume in section \ref{conclusion}.

\section{Calculation of 
``Goal quantity''
Illustrated with the Standard 
Model Group 
$S(U(2)\times U(3))$} 
\label{example}
Rather than going into using the 
structure as a {\em group} rather than
only the Lie-{\em algebra} structure 
we just above remarked that we 
can  determine the ``Abelian factor''
(see \ref{abelianf}) by studying the 
system of representation allowed as
representations of the {\em group}
rather than being just allowed by the 
Lie-algebra.

For example the phenomenological feature 
of the Standard Model, that gives rise
to, that the Standard Model Group indeed 
must be taken as $S(U(2)\times U(3))$ 
\cite{ORaifeartaigh}, is the restriction 
on the weak hypercharge $y$ quantization 
(or rather 
we prefer to use the half weak hypercharge 
$y/2$) 
realizing the usual 
assumption in the Standard Model about 
electric charge quantization (Milikan 
quantization extended with the well known 
rules for quarks). This rule become 
written for the Standard Model:
\begin{equation}
y/2 + I_W + ``triality''/3 = 0 (mod 1).
\label{restriction}
\end{equation}  

According to the rule to calculate the 
$\hbox{ Abelian factor }$ we shall in the 
limit of a going to infinity big range 
of $y/2$-values ask for what fraction 
of the number of values possible with 
the rule 
(\ref{restriction}) imposed and 
the same but only including 
representations with the simple groups 
$SU(2)$ and $SU(3)$ in the Lie algebra 
of the Standard Model
represented trivially.
If
we only allowed the adjoint or the trivial
representations of these simple groups,
so that $I_W=0(mod \, 1)$ and 
$``triality'' =0$, it is quite obvious 
in our Standard Model example, that 
the Standard Model rule (\ref{restriction})
allows, when the simple  representations
can be adjusted, all $y/2$ being an 
integer multiplum of 1/6. If we, however, 
limit the simple groups to have trivial
(or adjoint) representations only, then 
we can only have $y/2$ being integer.
It is clear that this means in the limit 
of  the large range $S$ that there are 
6 times
as many $y/2$ values allowed, when the 
representations of the simple groups 
are free, as when it is restricted to be
trivial (or adjoint). We therefore 
immediately find for the Standard Model 
Group 
\begin{equation}
\hbox{ ``Abelian factor''}_{S(U(2)\times U(3))} = 6^2 = 36.
\end{equation} 

In order to calculate the factor
``Semisimple factor '' 
(\ref{semisimplef}) we must look up 
the table for the $C_A/C_F$ for the simple
groups involved, then raise these factors 
to the power of the dimension of the
Lie-algebras in question, and very 
finally after 
having multiplied also by the 
``Abelian factor'' we must take the root 
of the total dimension of the whole 
group.  

\subsection{Useful Table}
Here we give the table to use, our 
(essentially inverse Dynkin index 
\cite{Dynkinindex})
ratios for the {\em simple} Lie groups,
with the representation $F$ selected so as 
to provide the biggest possible ratio 
$C_A/C_F$ still keeping $F$ non trivial, 
or let us say faithful (in a few cases 
the choice of this $F$ is not clear at 
the 
outset and the user of the table has 
to choose the largest number among
``vector'' and ``spinor'' after he has 
provided the rank $n$ he wants to use) :
    
{\bf Our Ratio of Adjoint to ``Simplest''
(or smallest)
Quadratic Casimirs $C_A/C_F$}
\begin{eqnarray}
\frac{C_A}{C_F}|_{A_n} & =&
\frac{2(n+1)^2}{n(n+2)} =
\frac{2(n+1)^2}{(n+1)^2 -1} =
\frac{2}{1-\frac{1}{(n+1)^2}}\\
\frac{C_A}{C_{F \; vector}}|_{B_n}& =&
\frac{2n-1}{n}= 2 - \frac{1}{n}\\
\frac{C_A}{C_{F \; spinor}}|_{B_n}&=
&\frac{2n-1}{\frac{2n^2 +n}{8}} =
\frac{16n -8}{n(2n+1)}\\
\frac{C_A}{C_F}|_{C_n} &=&
\frac{n+1}{n/2 +1/4} =
\frac{4(n+1)}{2n+1}\\
\frac{C_A}{C_{F \; vector}}|_{D_n}&=&
\frac{2(n-1)}{n-1/2}=
\frac{4(n-1)}{2n-1}\\
\frac{C_A}{C_{F \; spinor}}|_{D_n}&=&
\frac{2(n-1)}{\frac{2n^2-n}{8}}
= \frac{16(n-1)}{n(2n-1)}\\
\frac{C_A}{C_F}|_{G_2} &=& \frac{4}{2} =2\\
\frac{C_A}{C_F}|_{F_4} &=& \frac{9}{6} =
 \frac{3}{2}\\
\frac{C_A}{C_F}|_{E_6} &=&
\frac{12}{\frac{26}{3}} = \frac{18}{13}\\
\frac{C_A}{C_F}|_{E_7}&=&
\frac{18}{\frac{57}{4}} = \frac{72}{57}
= \frac{24}{19}\\
\frac{C_A}{C_F}|_{E_8}&=& \frac{30}{30} =1
\label{ratiotable}
\end{eqnarray}
For calculation of this table seek help
in\cite{Rittenberg,MacFarlaine}.

In the just above table we have of 
course used the conventional 
notation for the classification of Lie 
algebras, wherein the 
index $n$ on the capital letter denotes 
the rank (the rank $n$ is the maximal 
number of mutually commuting 
basis-vectors in the Lie algebra)  of the Lie algebra, and:
\begin{itemize}
\item $A_n$ is $SU(n+1)$,
\item $B_n$ is the odd dimension 
orthogonal group  Lie 
algebra i.e. for $SO(2n+1)$ or for its 
covering group $Spin(2n+1)$,
\item $C_n$ are the symplectic Lie 
algebras.
\item $D_n$ is the even dimension 
orthogonal Lie algebra i.e. for $SO(2n)$ 
or 
its covering group $Spin(2n)$,
\item while $F_4$, $G_2$, and $E_n$ for 
$n=6,7,8$ are the exceptional Lie 
algebras.  
\end{itemize}

The words $spinor$ or $vector$ following 
in the index the letter $F$,
which itself denotes the ``small'' 
representation - i.e. most promising 
for giving a small quadratic Casimir 
$C_F$ - means that we have 
used for $F$ respectively the smallest 
spinor and the smallest vector 
representation. 

\subsection{End of calculation of the
``goal quantity'' for the 
Standard Model 
Group}
 
Since the Lie-algebra in addition to the 
Abelian part ($U(1)$ usually called)
consists of $SU(2)$ and $SU(3)$ we must 
look these two simple Lie algebras up in
the table above, finding respectively
for the $C_A/C_F$ ratios
8/3 and 9/4, which must be taken to 
respectively the powers 3 and 8, since 
the dimensions of the $A_n =SU(n+1)$
Lie-groups are 
$``dimension'' = (n+1)^2-1$,
leading to 
\begin{eqnarray}
\hbox{``Semisimple factor''}_{S(U(")\times
U(3))} &=& (\frac{8}{3})^3 \cdot 
(\frac{9}{4})^8 = 3^{13}\cdot 2^{-7}=1594323/ 128\nonumber\\
 &=& 12455.6484375. 
\end{eqnarray}
Remembering that we got $6=3\cdot 2$ for 
the ratio of numbers of $y/2$-values,
 when all representation obeying 
(\ref{restriction}) were counted
relative to this number for only the 
representations with 
trivial 
representations of $SU(2)$ and $SU(3)$,
  the 
``Abelian factor''$= 6^2 = 3^2 \cdot 2^2$.
Then the whole factor, of which to next 
take the
12th root (since the total dimensionality
of the Standard Model group is 12) becomes
\begin{eqnarray}
``Semisimple factor'' \cdot 
``Abelian factor'' &=& 
(\frac{8}{3})^3(\frac{9}{4})^8 
\cdot
36 = 2^{-5} \cdot 3^{21} \nonumber\\
&=&448403.34375.  
\end{eqnarray}

Thus we just have to take the 12th root
of this quantity to obtain the score 
or ``goal quantity'' for the Standard 
Model {\em group} $S(U(2)\times U(3))$
\begin{eqnarray}
``goal quantity''_{S(U(2)\times U(3))}&=&
(2^{-5}\cdot 3^{15})^{1/12} = 3 \cdot 
(\frac{27}{32})^{1/12} =
3 \cdot 0.985941504 \nonumber\\
&=&{\bf 2.957824511}. 
\end{eqnarray}

Similar calculations give the 
``goal quantity'' for other groups.
But it requires of course either a lot 
of work or some rules and experiences 
with calculating such goal quantities
in order to see, which alternative groups
are the severe competitors of the 
Standard Model {\em group} $S(U(2)\times
U(3))$ that have to have their ``goal 
quantities'' computed in order to 
establish that the by Nature selected 
Standard Model group $S(U(2)\times
U(3))$ is indeed the winner 
in obtaining the highest ``goal quantity''
(except for groups being higher powers 
of the Standard Model group itself).

For example a very near competitor
is the group $U(2)$, for which one 
easily calculates
\begin{eqnarray}
``Semisimple factor''_{U(2)}& = & 
(\frac{8}{3})^3
= 2^{9}/3^3 = 18.962962963\\
``Abelian factor''_{U(2)}& =& 2^2 = 4\\
``goal quantity''_{U(2)}&=& 
(2^{11}/3^3)^{1/4}= 2^3/3 \cdot (\frac{3}{2})^
{1/4}\nonumber\\
 &=& 2^3/3 \cdot 1.10668192
 ={\bf 2.951151786}
\end{eqnarray}    
   
On the Fig.~\ref{fig1} we illustrate 
the three groups getting the three 
highest ``goal quantities''. The third 
group winning so to speak the bronze 
medal in this competition is $Spin(5)
\times SU(3)\times U(1)/ ``Z_6''$
(where $``Z_6''$ stands for a certain
with the integers modulo 6 isomorphic 
subgroup of
the center of the cross product group;
it arranges a quantization rule for 
the allowed representations quite 
analogous to that of the Standard 
Model group except, that the weak 
Lie algebra $SU(2)$ has been replaced 
by $Spin(5)$ (which is the covering 
group of $SO(5)$), which is very 
analogous to the Standard Model 
group just with SO(5) or rather 
Spin(5) which is its covering group
replacing the SU(2) in the Standard 
Model:

\begin{figure}       
\begin{center}
\includegraphics{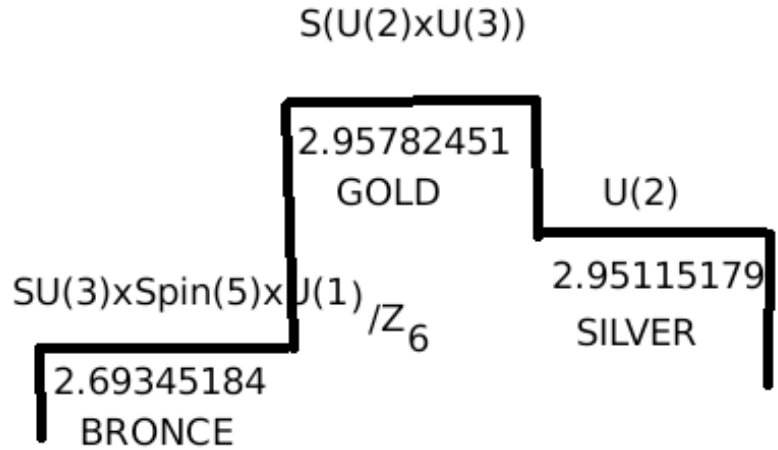}
\caption{\label{fig1}
This figure illustrates the three
Lie groups getting in our game the 
highest scores for our ``goal quantity''
as were the sportsmen winning gold silver 
and bronze medals.} 
  \end{center}
\end{figure}

\section{ Dimension of Space-time
Also} \label{dim}
The main point of my progress since last 
year \cite{seeking} is to say:

{\bf The choice of dimensionality of 
space time, that nature have made,} -
at least 3+1 for practical purpose -
{\bf can be considered also a choice
of a group,} - and even a gauge group, 
if we invoke general relativity -{\bf
namely say the Lorentz group or the 
Poincare group.} So if we have a 
``game'' or
a ``goal quantity'' selecting by letting
it be maximal the gauge group of the 
Standard Model, it is in principle 
possible to ask: 

{\bf Which among the as Lorentz or 
Poincare group applicable groups get 
the highest ``goal quantity'' score? 
Which dimension wins the competition 
among Lorentz or Poincare groups?.}

We would of course by extrapolation 
from the gauge group story (= previous 
work(with Don)\cite{seeking} ) expect 
that Nature should
again have chosen the ``winner''.

It is my point now that - with only very
little ``cheat'' - I can claim that indeed
{\bf  Nature has chosen 
that dimension $d=4$} (presumably meant 
to be the practical one, we see, and not 
necessarily the fundamental dimension, 
since 
our quantity could represent some 
stability against collapsing the 
dimension){\bf  that gives
the biggest score for the Poincare 
group!  } (for the Lorentz groups 
$d=4$ and $d=3$ share the winner 
place !)  

\subsection{  Development of Goal 
Quantities for dimension fitting.} 
\label{dgq}
In the present article  we shall ignore 
 anthropic principle arguments for what 
space time dimension should be, 
and seek to get a statement, that the 
experimental number of dimensions (4 if 
you count the truly observed one and 
take the convention to include time as
one dimension) 
just maximizes some quantity, that is a 
relatively simple function of the 
group structure of, say, the Lorentz 
group, 
and  which we then call
a ``goal quantity''.

\begin{center}
{\bf Making a ``goal quantity'' for 
Dimension is a Two step Procedure:}
\end{center}
\begin{itemize}
\item{ 1)} We 
first use the proposals 
in my work with Don Bennett 
 to give 
a number - a goal quantity -
for any Lie group. 
\item{2)} We have to specify 
on which group we shall take and use 
the procedure of the previous work; shall 
it be  the Lorentz group?, its covering 
group ?
or somehow an attempt with the Poincare 
group ?  :
\end{itemize}
\begin{center}
{\bf Developing a ``Goal quantity'' for 
``predicting''(fitting) the Space Time 
dimension}  
\end{center}
{A series of four proposals:}
\begin{itemize}
\item{a.} Just take the Lorentz group 
and calculate for that the inverse Dynkin 
index
or rather the quantity which we already 
used as ``goal quantity'' in the previous 
work and above (\ref{goal})
 $C_A/C_F$.
(Semi-simple Lorentz groups   
except for dimension $d=2$ or smaller
and in fact simple for 3 , 5 and higher).

\item{b.} We supplement in a somewhat 
{\em ad hoc way} the above {\em a.}, i.e. 
$C_A/C_F$ by taking its 
$\frac{d+1}{d-1}$th power. The idea 
behind this proposal 
is that we think of the 
{\em Poincare group }
instead of as under {\em a.} only on the
Lorentz group part, though still {\em in 
a crude way}. This means we think of 
a group, which is the Poincare group, 
except
that we for simplicity {\em ignore that 
the 
translation generators do not commute with
the Lorentz group part}. Then we assign
in accordance with the ad hoc rule
used for the gauge group the Abelian 
sub-Lie-algebra a formal replacement $1$
for the ratio of the quadratic Casimirs 
$C_A/C_f$
- because there is no limit to how small
momenta can be quantized and no natural 
way to obtain the charges $e_r$ for 
restricted representations, since we have 
essentially ${\bf R}$ as the Abelian 
group rather than $U(1)$ or complicated 
discrete subgroups $D$ being divided out -: 
I.e. we put  ``$e_A^2/e_F^2$'' = 
$``C_A/C_F|_{Abelean \ formal}''
= 1$. Next we construct an ``average''
averaged {\em in a logarithmic way} 
(meaning 
that we average the logarithms and then 
exponentiate again) weighted with the 
dimension of the Lie groups over all the 
dimensions of the Poincare Lie group.
Since the dimension of the Lorentz group 
for $d$ dimensional space-time is 
$\frac{d(d-1)}{2}$ while the Poincare 
group
has dimension$\frac{d(d-1)}{2} +d = 
\frac{d(d+1)}{2}$ the logarithmic 
averaging means that we get
\begin{eqnarray}
exp( \frac{\frac{d(d-1)}{2} 
\ln(C_A/C_F)|_{Lorentz} + \ln(1)*d}
{d(d+1)/2})
&=& (C_A/C_F)|_{Lorentz}^{\frac{d(d-1)}{2}
/ \frac{d(d+1)}{2}} \nonumber\\
&=& (C_A/C_F)|_{Lorentz}
^{\frac{d-1}{d+1}}\label{f2} 
\end{eqnarray} 
  
That is to say we shall make a 
certain ad hoc partial inclusion 
of the Abelian dimensions in the 
Poincare groups. 

To be concrete we here propose to 
say crudely: Let the Poincare 
group have of course $d$ ``Abelian'' 
generators or dimensions. Let the 
dimension of the Lorentz group 
be $d_{Lor} = d(d-1)/2$; then the 
total dimension of the Poincare group 
is $d_{Poi} = d + d_{Lor} = d(d+1)/2$.
If we crudely followed the idea 
of weighting proposed in the previous 
article \cite{seeking} or above 
(\ref{goal} as if the 
$d$ ``abelian'' generators were 
just simple cross product factors 
- and not as they really are:
not quite usual, because they do not 
commute 
with the Lorentz generators - then 
since we formally are from this 
previous article suggested to 
use the {\em as if number 1 for the 
Abelian 
groups}, we should use the quantity
\begin{equation}
(C_A/C_F)|_{Lor}^{\frac{d_{Lor}}{d_{Poi}}}
= (C_A/C_F)|_{Lor}^{\frac{d-1}{d+1}}
\end{equation}  
 as goal quantity.

Really you can simply say: we put 
the ``Abelian factor '' =1, but still 
take the $d_{Poi}= d(d+1)/2$th  root
at the end, by using the total 
dimension of the Poincare group
$d_{Poi}$. The crux of taking this 
``$1$'' is that we do not have anything 
corresponding to the division out of a 
discrete group giving the restriction like
(\ref{restriction} in the Poincare case.   
\item{c.} We could improve the above 
proposals for goal quantities {\em a.}
or {\em b.} by including into the 
quadratic Casimir $C_A$ for the adjoint 
representation also contributions from 
the translation generating  generators,
so as to define a quadratic Casimir for 
the whole Poincare group. This would 
mean, that we for calculating our goal
quantity would do as above but 
\begin{equation}
{\bf Replace:} C_A \rightarrow C_A +C_V,
\end{equation}
where $C_V$ is the vector representation 
quadratic Casimir, meaning the 
representation under which the translation
generators transform under the Lorentz 
group. Since in the below table we in 
the lines denoted ``no fermions'' have
taken the ``small representation'' $F$
to be this vector representation $V$,
this replacement means, that we replace 
the
goal quantity ratio $C_A/C_F$ like this:
\begin{eqnarray}
\hbox{{\bf  (S)O(d),}}&& 
\hbox{{\bf ``no spinors'':}}\nonumber\\
C_A/C_F= C_A/C_V &&\rightarrow  
(C_A+C_V)/C_F = C_A/C_F + 1\\
\hbox{{\bf Spin(d),}}&&
\hbox{{{\bf ``with spinors''}}:}\nonumber\\
C_A/C_F&&\rightarrow  (C_A+C_V)/C_F\nonumber\\
&&=
C_A/C_F + (C_A/C_V)^{-1}(C_A/C_F)\nonumber\\
& &= 
(1 + 
(C_A/C_F)|_{\hbox{no spinors}}^{-1})C_A/C_F.   
\end{eqnarray}

{\em Let me stress though that this 
proposal c. is not quite ``fair'' in as 
far as it is based on the Poincare group,
while the representations considered 
are \underline{not} faithful w.r.t. 
to the whole Poincare group, but only 
w.r.t. the Lorentz group} 


\item{d.} To make the proposal {\em c.}
a bit more ``fair'' we should at least 
say: Since we in {\em c.} considered a 
representation which were only faithful
w.r.t. the Lorentz subgroup of the 
Poincare group we should at least correct
the quadratic Casimir - expected crudely
to be ``proportional'' to the number of 
dimensions of the (Lie)group - by a factor
$\frac{d+1}{d-1}$ being the ratio of the 
dimension of the 
Poincare (Lie)group, $d + d(d-1)/2$ 
to that of actually faithfully represented
Lorentz group $d(d-1)/2$. That is to say
we should before forming the ratio of the
improved $C_A$ meaning $C_A+C_V$ (as 
calculated under {\em c.}) to $ C_F$ 
replace this $C_F$ by 
$\frac{d+1}{d-1}*C_F$, i.e. we perform 
the replacement:
\begin{equation}
C_F \rightarrow C_F*\frac{d(d-1)/2 +d}{
d(d-2)/2} = C_F*\frac{d+1}{d-1}.
\end{equation}
        
Inserted into $(C_A+C_V)/C_F$ from 
{\em c.}
we obtain for the in this way made more
``fair'' approximate ``goal quantity''
\begin{eqnarray}
\hbox{``goal quantity''}|_
{\hbox{no spinor}} &=&(C_A/C_F +1)*
\frac{d-1}{d+1}\\
\hbox{``goal quantity''}|_
{\hbox{w. spinor}} &=&( 1+ 
(C_A/C_F)|_{\hbox{no spinor}}^{-1})*C_A/C_F*
\frac{d-1}{d+1}\nonumber\\
\end{eqnarray}
{\em This proposal {\em d.} should then 
at least be crudely balanced with respect
to how many dimensions that are 
represented faithfully.}
\end{itemize}  

\subsection{Philosophy of the goal
quantity 
construction/development}   
The reader should consider these different
proposals for a quantity to maximize
(= use as goal quantity) as rather closely 
related versions of a quantity suggested 
by a perhaps a bit vague ideas being 
improved successively by treating the 
from our point of view a bit more 
difficult to treat Abelian part (=the 
translation part of the Poincare group)
at least in an approximate way. 

One should have in mind, that this 
somewhat
vague basic idea behind is: The group 
selected by nature is the one that 
counted
in a ``normalization determined from the 
Lie algebra of the group'' can be said to 
have a faithful representation ($F$)
the matrices of which move as little as 
possible, when the group element being 
represented move around in the group.

Let me at least clarify a bit, what is 
meant by this statement:

We think by representations as usual
on linear representations, and thus it 
really means representation of the group 
by means of a homomorphism of the group
into a group of matrices. The requirement 
of the representation being faithful then 
means, that this group of matrices shall
actually be an isomorphic image of the 
original group. Now on a system of 
matrices we have a natural metric, 
namely the metric in which the distance 
between two matrices ${\bf A}$ and 
${\bf B}$ is given by the square root of 
the trace of the numerical square of the 
difference
\begin{equation}
dist = \sqrt{tr(({\bf A -B})
({\bf A -B})^+)}. \label{distrep}
\end{equation}  
To make a comparison of one group and 
some representation of it  with 
another group and its representation 
w.r.t. to, how fast the representation 
matrices 
move for a given motion of the group
elements, we need a normalization giving 
us a well-defined metric on the groups, 
w.r.t. which we can ask for the rate 
of variation of the representations.
In my short statement I suggested 
that this ``normalization should be 
determined from the Lie algebra of the 
group''. This is to be taken to mean 
more precisely, that one shall consider 
the {\em adjoint} representation, which
is in fact completely given by the 
Lie algebra, and then use the same 
distance concept as we just proposed for 
the matrix representation 
$ \sqrt{tr(({\bf A -B})
({\bf A -B})^+)}$. In this way the 
quantity to minimize would be the ratio
of the motion-distance in the 
representation - $F$ say - and in the 
Lie algebra representation - i.e. the 
adjoint representation. But that ratio 
is just for infinitesimal motions 
$\sqrt{C_F/C_A}$. So if we instead 
of talking about what to minimize, 
inverted it and claimed we should 
maximize we would get $\sqrt{C_A/C_F}$ 
to be {\em 
maximized}. Of course the square root 
does not matter, and we thus obtain in 
this 
way a means to look at the ratio $C_A/C_F$
as a measure for the motion of an element 
in the group
compared to the same element motion on 
the representation.

It might not really be so wild to think
that a group which can be represented in 
a way so that the representation varies 
little when the group element moves around
would be easier to get realized in nature 
than one that varies more. If one imagine 
that the potential groups become  good 
symmetries by 
accident, then at least it would be less 
of an accident required the less the 
degrees of 
freedom moves around under the to the 
group corresponding 
  symmetry (approximately). It is really 
such a 
philosophy of it being easier to get some 
groups approximately being  good 
symmetries
than other, and those with biggest 
$C_A/C_F$ should be the easiest to  
become
good symmetries by accident, we argue 
for. That is indeed the 
speculation behind the present article
as well as the previous one 
\cite{seeking} that symmetries may 
appear 
by accident(then perhaps being 
strengthened to be exact by some means
\cite{Foerster,Damgaard}).

\subsection{Phenomenological 
Philosophy}\label{phen}
But let us stress that you can also look 
at the present work and the previous one 
in the following phenomenological 
philosophy:

We wonder, why Nature has chosen just 4
(=3+1) 
dimensions and why Nature - at the 
present 
experimentally accessible scale at 
least - 
has chosen just the Standard Model 
group $S(U(2)\times U(3))$? Then we 
speculate that there might be some 
quantity characterizing groups, which 
measures how well they ``are suited '' 
to be the groups for Nature. And then we 
begin to {\em seek} that quantity as 
being some 
function defined on the class of abstract 
groups - i.e. 
giving a number for each abstract (Lie?) 
group -
of course by proposing for ourselves at 
least  various versions or ideas 
for what such a {\em relatively simple}
function defined on the abstract Lie 
groups could be. Then the present works 
- this paper and the previous 
ones\cite{seeking} and \cite{dimension} - 
represents the present status of the 
search: We found that with small 
variations the types of such functions
representing the spirit of the 
{\em little 
motion of the ``best'' faithful 
representation},i.e. essentially the 
largest $C_A/C_F$, 
turned out truly to bring Natures choices 
to be (essentially) the winners.

In this sense we may then claim that we 
have found by phenomenology, that at 
least the ``direction'' of a quantity like 
$C_A/C_F$ or light modifications of it 
is a very good quantity to make up a 
``theory'' for, why we have got the 
groups we got!

Here we bring the table in which we 
present the calculations of our for the 
space-time dimension relevant various 
``goal quantities'':

\begin{center}
\hspace{-3mm} \begin{tabular}{ | l | l | 
l | l 
|l|l|l|}
\hline
Di-
& Lorentz 
& Ratio 
&Ratio 
& 
{\em c.}-quan-
& $\frac{d-1}{d+1}$
& {\em d.}-quan-
\\
men-  
& group,  &$C_A/C_F$   
& $C_A/C_F$ as  
&
tity
& &
tity
\\
sion&covering&for spinor&no spinor&
max c)&&max d)\\
\hline
2\footnote{the treatment of the case d=2 becomes because of the 
Abelianness of Lorentz group ill defined, and we have to use the somewhat formal specification from earlier work
 for the ``with spinor'' we even use 
formally the special rule connected with dividing out a subgroup of the center,
but since d=2 gets beaten by the higher dimensions the details are not so serious for us.}  & U(1)& -(for-
& -(for-
&4&1/3&4/3
\\
&& mally 2)& mally 1)&&&=1.33\\ 
\hline
3& spin(3) & $\frac{8}{3} = 2.67$&1&
$\frac{16}{3}=5.3$ 
&$ \frac{2}{4}$& 
$\frac{8}{3}= 2.67$ \\
\hline
4&$Spin(4)$  &$\frac{8}
{3}$
 &$\frac{4}{3}$& $\frac{14}{3}$
 &$\frac{3}{5} $ &$\frac{14}{5}$
\\
&$SU(2)\times$ 
&=2.67  &&=4.67  &&\\
&$SU(2)$&&&&&=2.8\\
\hline
 5   & Spin(5) &$\frac{12}{5}=2.4 $&$\frac
{3}{2} =1.5$&
$4$  
&$\frac{4}{6}$&$\frac{8}{3}= 2.67$  
\\
\hline
6 &$Spin(6)$&$\frac{32}{15}$&$\frac{8}{5}
=1.6$
&$\frac{52}{15}=3.5 $  &
$\frac{5}{7}$&
$\frac{52}{21} =2.5$  \\
\hline
d 
& Spin(d)& 
$\frac{8(2n-1)}{n(2n+1)}=$&$2- 1/n=$   &
$\frac{8(3d-5)}{d(d-1)}$&$\frac{d-1}{d+1}$&
$\frac{8(3d-5)}{d(d+1)}$\\
odd&&$\frac{16(d-2)}{d(d-1)}$&$2-
\frac{2}{d-1}$ &&&\\
\hline
d 
& $Spin(d)$ & $ \frac{16(
d-2)}{d(d-1)}$&$\frac{4(n-1)}{2n-1}
$ &
$\frac{8(3d-5)}{d(d-1)}$ &
$\frac{d-1}{d+1}$
&$\frac{8(3d-5)}{d(d+1)}$ \\ 
even&&&$=\frac{2d-4}{d-1}$    &&&\\
\hline
d  
& Spin(d) & 
 $\approx 16/d$ &$\rightarrow 2$    
& $\approx 24/d$&$\rightarrow 1$& 
$\approx 24/d 
$\\
$ \infty$&&&&&&$\rightarrow 0$\\
\hline
d 
& Spin(d) & 
 $\approx 16/d$ &$\rightarrow 2$    
& $\approx 24/d$&$\rightarrow 1$& 
$\approx 
24/d$ 
\\
$ \infty$&&&&&&$\rightarrow 0$\\
\hline
\hline

\end{tabular} 
\end{center}

{\bf Caption:}
We have  put the goal-numbers for the 
third proposal {\em c} in which I (a bit 
more in detail) seek to make an analogon to 
the number used in the reference 
\cite{seeking} in which we studied the
gauge group of the Standard Model.
The purpose of {\em c.} is to approximate 
using  the {\em Poincare 
group} a bit more detailed, but still 
not by making a true representation 
of the Poincare group. I.e. it is still 
not truly the Poincare group we represent 
faithfully, but only the Lorentz group,
or here in the table only the covering 
group $Spin(d)$ of the Lorentz group.
However, I include in the column 
marked ``{\em c.}, max c)'' in the 
quadratic Casimir $C_A$ of the Lorentz 
group an extra term coming from the 
structure constants describing the 
non-commutativity of the Lorentz group 
generators with the translation generators
$C_V$ so as to replace $C_A$ in the 
starting expression of ours $C_A/C_F$
by $C_A + C_V$. In the column marked 
``{\em d.}, max d) '' we correct the ratio
to be more ``fair'' by counting at least 
that because of truly faithfully 
represented part of the Poincare group 
in the representations, I use, has only 
dimension $d(d-1)/2$ (it is namely 
only the Lorentz group) while the full
Poincare group - which were already in 
{\em c.} but also in {\em d.} used in 
the improved $C_A$ being $C_A +C_V$ -
is $d(d-1)/2 + d = d(d+1)/2$. The 
correction is crudely  made by the 
dimension ratio $dim(Lorentz)/dim(Poincare)
=(d-1)/(d+1)$ given in the next to last
column.  
\begin{center}
 \begin{tabular}{ | l | l | l 
|l|l|l|}
\hline
Di-
& Lorentz 
& Ratio 
&Ratio 
& 
{\em c.}-
& 
{\em d.}-
\\
men-&group&$C_F/C_A$&$C_A/C_F$&quantity&
quantity\\
sion  &(covering)  &for spinor   
& ``no spinor'' 
&max c)& 
max d)\\
\hline
2\footnote{the treatment of the case d=2 becomes because of the 
Abelianness of Lorentz group illdefined, and we have to use the somewhat formal specification from \cite{seeking}; for the ``with spinor'' we even use 
formally the special rule connected with dividing out a subgroup of the center,
but since d=2 gets beaten by the higher dimensions the details are not so serious for us.}  & U(1)& -(f.: 2)& -(f.: 1)&4&
4/3=1.33\\ 
\hline
3& spin(3) & $\frac{8}{3} = 2.67$&1&
$\frac{16}{3}=5.33$ 
&
$\frac{8}{3}= 2.67$ \\
\hline
4&$Spin(4)$  &$\frac{8}
{3}= 2.67$ &$\frac{4}{3}$& $\frac{14}{3}
 =4.67$ &
$\frac{14}{5}
= 2.8$ \\
&$=SU(2)$
&&&&
\\
&$\times SU(2)$&&&&
\\
\hline
 5   & Spin(5) &$\frac{12}{5}=2.4 $&$\frac
{3}{2} =1.5$&
$4$  
&
$\frac{8}{3}= 2.6667$  
\\
\hline
6 &$Spin(6)$&$\frac{32}{15}$&$\frac{8}{5}
=1.6$
&$\frac{52}{15}=3.47 $  &
$\frac{52}{21} =2.4762$  \\
\hline
d odd & Spin(d)& 
$\frac{8(2n-1)}{n(2n+1)}$&$2- 1/n=$   &
$\frac{8(3d-5)}{d(d-1)}$&
$\frac{8(3d-5)}{d(d+1)}$\\
&&$=\frac{16(d-2)}{d(d-1)}$&$2-2/(d-1)$ &
&\\
\hline
d even& $Spin(d)$ & $ \frac{16(
d-2)}{d(d-1)}$&$\frac{4(n-1)}{2n-1}=\frac{
2d-4}{d-1}$ &
$\frac{8(3d-5)}{d(d-1)}$ 
&$\frac{8(3d-5)}{d(d+1)}$ \\ 
\hline
d odd 
& Spin(d) & 
 $\approx 16/d$ &$\rightarrow 2$    
& $\approx 24/d$&
$\approx 24/d \rightarrow 0$\\
$\rightarrow \infty$&&&&&
\\
\hline
d even 
& Spin(d) & 
 $\approx 16/d$ &$\rightarrow 2$    
& $\approx 24/d$&
$\approx 
24/d \rightarrow 0$\\
$\rightarrow \infty$&&&&&\\
\hline
\hline

\end{tabular} 
\end{center}



\section{The Higgs Representation}
\label{Higgs}
A rather simple and successful application
of our ideas is to seek the answer to the
question: Why has the Higgs field just 
got the representation $({\bf 2} , {\bf 1},
y/2 = 1/2)$ under the Standard Model
group with the Lie algebra factors 
written in the order $SU(2)\times SU(3)
\times U(1)$ ? 

Note that the selection of the gauge 
group by our ``goal quantity'' had 
the character of being obtained as 
a ratio - of the quadratic Casimirs 
$C_A$ for adjoint and $C_F$ for another 
faithful representation or some 
``replacements'' for them in the Abelian
cases - of an adjoint representation 
parameter to one for {\em another 
representation}$F$. Also this other 
representation $F$ gets basically selected
by the same principle as the selection
of the whole gauge group by maximizing 
our ``goal quantity'', because we also 
{\em select the representation 
$F$ from the requirement that our 
``goal quantity '' be maximized}.

Thus in reality we have hit on a quantity
that tends to select both a {\em group}
and a smallest $C_F$ representation.

Now strictly speaking most irreducible 
representations of say the Standard Model
group $S(U(2)\times U(3))$ will not 
usually be completely faithful. It is 
rather so that the various representations
$F$ appearing as representations of the 
simple subgroups will not be truly 
faithful, but rather {\em only be faithful 
for some subgroup} of the  
$S(U(2)\times U(3))$ group say. 
If we therefore now shall make some 
numbers assigned to the various{\em not 
completely faithful} representations 
which are allowed as representations 
of the Standard Model {\em group} 
$S(U(2)\times U(3))$, it would be most 
``fair'' to count the ratio of the 
quadratic Casimir in the ``Adjoint''
representation - or better in the 
group itself - {\em by not using the
full say Standard Model Group, but rather
only that part of the group 
$S(U(2)\times U(3))$ that is indeed 
faithfully represented on the 
representation, which is up to be tested,
with a number to specify which 
representation should be favoured}.

So let us say we have some representation
$R$ of say the Standard Model 
{\em group}, i.e.
an allowed one, which of course then also 
obeys the quantization rule (such as) 
(\ref{restriction}).

Now there is always a kernel $K$ 
consisting 
of the elements in the group 
$S(U(2)\times U(3))$ or more generally
the Lie group $G$, with which we work,
for which the elements in $R$ are 
transformed trivialy, it means not
shifting to another element, but only to 
itself. This kernel $K$ is of course 
an invariant subgroup of the full group
$G$. This means that $G/K$ is a well defined
factor group in $G$. Then we should 
naturally suggest the ``fair'' rule 
that we construct the number according to
 which the representation $R$ should be 
selected {\em as the number we would 
get by calculating the ``goal quantity ''
of ours for the group $G/R$ with though 
the restriction that the $F$ should 
correspond to $R$.}

Let us illustrate this rule proposed 
by looking at a couple of examples:

If we want to consider one of the 
representations $F$ giving the 
maximal $G_A/G_F$ for one of the 
simple subgroups, which in Standard Model
can only be $SU(2)$ or $SU(3)$, then 
for these
two groups the $F$-representations 
are respectively ${\bf 2}$ and ${\bf 3}$
(or one could take the equivalent ${\bf 
\bar{3}}$ for the $SU(3)$). But of course
say ${\bf 2}$ alone without any $y/2$ 
charge would not be allowed by the 
Standard Model group $S(U(2)\times U(3))$.
Thus we are forced to include an 
appropriate $y/2$. Doing that you 
can easily find that the relevant 
factor groups $S(U(2)\times U(3))/K$
becomes in the two cases respectively
$U(2)$ and $U(3)$. Actually with 
smallest $y/2$ values allowed in the two
cases $y/2 = 1/2 $ ( or $-1/2$) for
$SU(2)$ and $y/2 = -1/3$ for $SU(3)$
with ${\bf 3}$ we get just the 
same $F$ as is used in our calculations
of our ``goal quantity''. This means 
that quantities to select the 
representation happens to be in our two 
cases just the ``goal quantities''
for the two groups $U(2)$ and $U(3)$,
namely just the factor groups.
We already know that $U(2)$ were the 
``silver medal winner'' and thus that 
it should be trivially  $ U(2)$
related to measuring the size of the 
representation 
${\bf 2}, {\bf 1}, y/2 =1/2)$ which gets 
selected. This means the winning 
- and that means ``smallest'' 
representation of the Standard Model 
(measured by using the associated factor 
group for which it is faithful) - 
representation of the Standard Model 
group became this 
${\bf 2}, {\bf 1}, y/2 =1/2)$.
 This is just the
representation of the Higgs. So the Higgs 
representation is predicted this way
(as the ``smallest'' in our way of 
counting, closely related to the game
we used to tell the gauge group with) !

\section{The (chiral) Fermion 
Representations}
\label{fermionrep}
It is now the idea to use the very same 
``goal quantity'' as the one, with which 
we exercised in deriving the Higgs 
representation above, to argue for the 
Fermion representations in the Standard
Model - or rather what we in the present 
philosophy expect for the  choice of 
Nature - as to what they should  be.

Here the situation is somewhat more 
complicated because the requirement 
that there be no gauge- nor gauge gravity 
anomalies imposes restrictions on the 
whole system of representations for the 
chiral fermions. Assuming that we work 
with 3+1 dimensions we can take it as 
our convention to work with only 
left handed spin 1/2 Fermions, because 
we can let the right handed ones simply
be represented by their CP-analogue 
left handed ones. 

We must therefore first write down 
the non-anomaly conditions for having 
various thinkable numbers of families 
for the various representations of the 
Standard Model. 

Now the use of anomaly conditions together
with the assumption of ``small 
representations'' (in some meaning or the 
other) 
we already used in some articles years 
ago. For instance in ``Why do we have 
parity violation?''\cite{parity} Colin Froggatt and I
sought to answer this question by using 
the principle of small representations 
to derive the representations that the 
Standard Model should have and thus why 
they would give parity violation the 
well known way. Also in \cite{CostaRica} 
we allude to the principle of small 
representations
(here in the last section). In fact in 
the section XIII, called ``Hahn-Nambu-like
Charges'' we sought to derive the system
of the representations of Weyl Fermions 
(we use a notation there of only counting 
the left handed spin 1/2 fermions, letting
it be understood that the right handed 
components achieved by CP i.e. of the anti
particles of course exists but are just 
not listed in the way we keep track 
of the particles in this notation; 
that is to say that normally considered
right handed Weyl particles are just 
counted by their CP-antiparticle, which 
if left handed). We sought to derive it 
from the no-anomaly-conditions and a 
principle of ``small representations''.
The latter were not exactly the same as 
we seek to develop in the present more 
recent but in some approximate sense
it were very close to the present idea 
of a small representation principle, as
we claim the choice of nature of the 
Standard Model group $S(U(2)\times
U(3))$ indicates. Nevertheless the two 
ideas of a ``small representation 
principle'' are so close that at least 
I give/gave them the same name ``small
representations''. 

In the section XIII of the Puerto-Rico
conference proceedings \cite{CostaRica}
we use somewhat special technology to 
argue that{\em  imposing the conditions for:'} 
\begin{itemize}
\item{1.} no
chiral anomalies and no mixed anomalies
for the gauge charge conservations,
\item{2.} 
together with a {\em small representation 
principle} (formulated using the concept 
to be explained of ``Hahn Nambu charges'')
\item{3.} and that the fermions shall 
be {\em mass protected} (i.e. get zero 
mass due to gauge (charge) conservation, 
were 
it not for the ``Higgsing''),
\end{itemize}
  {\em lead to the Standard Model 
representations spectrum} basically
(i.e. we get that there should be a number
of families of the type we know, but how 
many we do not predict from these 
assumptions). 
 
The technology used in \cite{CostaRica}  
was to consider only 
a certain subset of charges - called 
there ``Hahn-Nambu charges'' - of the 
Cartan 
algebra of the Standard Model Lie 
algebra, or of the Lie algebra for 
any other gauge grouop being discussed.

Since the rank of the Standard model 
(gauge)group is 4, there are of course 4 
linearly independent Cartan algebra 
charges. But now we used in the reference
\cite{CostaRica} not  linearly 
independent charges, but rather linear
combinations of the Cartan algebra charges
selected to have the special property,
that for representations allowed for the 
Standard Model {\em group} these specially
selected Cartan algebra charges      
had only the integer values and even in 
the usual Standard Model system of 
representaions took only the values 
$-1, 0, \hbox{or} 1$. 

Let me explain the technique of our 
Costa Rica proceedings paper 
\cite{CostaRica} a bit more:

Starting from assuming a gauge group
with rank four say (but we really have 
in mind using a similar discussion on any
potential gauge group, so that we also 
with those considerations could hope for 
approaching a derivation of 
an answer to why just the Standard Model) 
and deciding to 
consider only the Cartan algebra part,
we would basically have assumed 
effectively an ${\bf R}^4$ gauge Lie 
algebra. But as a rudiment of as well the 
explicite charge quantization rule 
resulting from the {\em group} structure
as from the charge quantizations caused 
by the non-abelian Lie algebra structure 
present before we threw the non-abelian 
parts away - only keeping the Cartan 
algebra - we would have quatization 
rules for the Cartan algebra charges.
Indeed we would rather obtain an effective
gauge {\em group} after this keeping 
nothing but the Cartan algebra being 
$U(1)^4$ than the here first mentioned 
${\bf R}^4$. This would mean that in the
appropriate basis choice for these Cartan
algebra charges they would all be 
restricted by the group structure to be 
integers. Making sums and/or differences 
of such ``basis'' charges restricted to be 
integers one can easily write down 
combinations which again would be 
restricted to have only integer charges.

But now the main question of interest 
in our earlier quantization of certain
Cartan algebra charges were to implement 
the requirement/assumption of ``small
representations'' or for Abelian 
equivalently ``small charges''. 

We formulated the requirement of 
such ``small charges'' via defining a 
concept of a ``Hahn Nambu charge''.
Such a type of charge, which we would 
denote as ``Hahn Nambu charge'', were
by our definition assumed to obey:
\begin{itemize}
\item A ``Hahn Nambu charge'' should 
be one of the combinations of the 
Cartan algebra charges, which precisely
were allowed to take on integers - no more
no less - (due to the group structure 
of the $U(1)^4$ say for rank 4).
\item But in the actual detailed model
one should for the Hahn Nambu charge 
{\em only find the charge eigenvalues 
$-1, 0, \hbox{or} 1$}. (This assumption 
is, one may  say, an assumption of small 
charge values -for the Hahn Nambu charge 
type - in as far as the charge value 
numerically less than or equal to 1 
is ``small'' compared to the quantization
interval assumed just above to be 1).  
\end{itemize}  

Then instead of assuming in some other 
way, that we seek a model with the smallest
possible charge values, we used then 
in \cite{CostaRica} and \cite{parity} to 
say in stead - and crudely 
equvalently - that we should arrange so 
many ``Hahn Nambu charges'' to exist in 
the model to be sought as possible. 

In order that the reader shall get an 
idea what 
type of charges these ``Hahn Nambu 
charges'' 
are, let me mention the Hahn Nambu 
charges
of the Standard Model:
\begin{eqnarray}
``HNred'' &=&y/2 + I_{W3} 
+\sqrt{3}\lambda_{8red}\\
 ``HNblue'' &=&y/2 + I_{W3} 
+\sqrt{3}\lambda_{8blue}\\
``HNyellow'' &=&y/2 + I_{W3} 
+\sqrt{3}\lambda_{8yellow}\\
\lambda_{2yellow}&=&\left ( 
\begin{tabular}{ccc}
1&0&0\\
0&-1&0\\
0&0&0\\
\end{tabular}\right )\\
\lambda_{2yellow}&=&\left ( 
\begin{tabular}{ccc}
1&0&0\\
0&-1&0\\
0&0&0\\
\end{tabular}\right )\\
\lambda_{2red}&=&\left ( 
\begin{tabular}{ccc}
0&0&0\\
0&1&0\\
0&0&-1\\
\end{tabular}\right )\\
\lambda_{2blue}&=&\left ( 
\begin{tabular}{ccc}
-1&0&0\\
0&0&0\\
0&0&1\\
\end{tabular}\right )\\
\hbox{``Twise weak isospin $I_{W3}$''}&=&2 I_{W3} 
\end{eqnarray} 

Here we have used 
 a notation, wherein the colors are 
listed in the series ($``red'$', $``blue''$, 
$``yellow''$) in columns and rows and 
defined  the variously 
colordefined 
$\lambda_8$-matrices:
\begin{eqnarray}
\sqrt{3}\lambda_{8red}&=&\left (\begin{tabular}{ccc}
-2/3&0&0\\
0&1/3&0\\
0&0&1/3
\end{tabular} \right )\\
\sqrt{3}\lambda_{8blue}&=&\left (\begin{tabular}{ccc}
1/3&0&0\\
0&-2/3&0\\
0&0&1/3
\end{tabular} \right )\\
\sqrt{3}\lambda_{8yellow}&=&\left 
(\begin{tabular}{ccc}
1/3&0&0\\
0&1/3&0\\
0&0&-2/3
\end{tabular} \right )
\end{eqnarray}
It is easy to check that these 7 
``Hahn Nambu charges'' are related to
each other by  being sums or 
differences of each other, and also that
they are indeed according to our 
definition indeed ``Hahn Nambu charges''
in the wellknown Standard Model. Indeed 
you should also see that the first 
three of them, ``HNred'', ``HNblue'',
and ``HNyellow'' are indeed three 
color choices for what historically 
Hahn and Nambu proposed as the electric 
charge to be used in a QCD including model.
Nowadays we know, that quarks only have 
electric charges $2/3$ or $-1/3$
fundamental charges, but 
the original Hahn Nambu charge were
precisely constructed to have only the 
integer Millikan charge {\em even for 
the quarks}.

The crux of the calculation, we want to 
extract from the study of Hahn Nambu 
charges in our old works \cite{CostaRica,parity}, is that imposing the no gauge 
anomaly conditions for the Cartan 
subalgebra, using the the assumption that
we have as many ``Hahn Nambu charges'' as
possible still having a mass protected 
system of (Weyl)fermions, we are led 
to a system of representations which 
{\em is} indeed the usual one when 
extended to get the non-abelian 
charges too.

The technique we used in the old paper(s)
\cite{CostaRica} were in fact to study 
the no-anomaly constraint equations 
moulo 2, which for Hahn Nambu charges,
that never take by assumption/definiton
charge values bigger than 1 numerically,
close to be enough. 

Actually it turned out that we could 
first find a system of mass-protected
Weyl-fermions, when the dimension of
the Cartan algebra (= the number of 
linearly independent Hahn Nambu charges)
became at least 4. In that case then 
we had indeed to have a system of 
Weyl fermions, which modulo some 
trivial symmetries, had to be the 
one found experimentally w.r.t.
these ``Hahn Nambu charges''.

This should be interpreted to say, 
that requiring maximal numbers of
``Hahn Nambu Charges'' in our sense,
which is a slightly special way of 
requiring small representations together
with the assumtions of mass protection and no anomalies, leads to the Standard Model
fermion system.

That is to say we should consider 
the structure of a family in the Standard
Model to essentially come out of such
requirements. In this way we can 
count the fermion system/spectrum 
as largely being a successful
result comming out from a 
``Small representation principle''!

\section{Speculations on the 
Full Group
of Gauge Transformations and 
Diffeomorphism Symmetry}
\label{full}
In the above discussion and in the 
previous articles in the present series 
of papers \cite{seeking,dimension} we
sought to find a game leading to the 
``gauge group''. But now we want to have 
in mind that the ``gauge group'' is not
truly the most physical and simple concept
in as far as the true symmetry in a 
gauge theory with ``gauge group''$G$ is 
really not truly $G$, but rather a cross 
product 
of one copy of $G$, say $G(x)$ for every 
point $x$ in space time. That is to 
say the true symmetry group of the 
gauge theory having the ``gauge group''
$G$ is rather $ {\hbox{{\huge\bf  
$\times$}}}_x G(x) =
G\times G \times \cdots \times G$,
where in the cross product it is supposed
that we have one factor for every space 
time point $x$.   

Above we saw that the goal quantity 
for a group were suggested to be of a type,
that is balanced in such a way, that 
the score or goal quantity is the {\em 
same for a group $G$ and for the cross 
product of this group with itself 
$G\times G \times \cdots \times G 
\times G$ any number of times.} 

This means of course, if as we found the 
Standard Model group $S(U(2)\times U(3))$
wins our game, then in fact any product 
of this group with itself any number of 
times can also be said to get just 
the same score, and thus it will also win!
That it to say that we might reinterpret 
our work by saying: It is not truly 
the gauge group for the realized gauge 
theory we predict to be the winner. Rather 
we could say that the group that wins 
is the whole symmetry group of the full
quantum field theory supposed to be 
realized. The concept of the full 
gauge symmetry (or we could say 
reformulation symmetry) is -- we would say
-- a simpler concept than the concept
of the ``gauge group'' for which it would 
have to be specified how this gauge 
group would have to be applied, namely 
one should construct a group of all 
gauge transformations   
 $ {\hbox{{\huge\bf  
$\times$}}}_x G(x) =
G\times G \times \cdots \times G$.

But since this full group gets just the 
same score as the more complicatedly 
defined ``gauge group'' we could claim 
that our prediction is, that it is this 
group of all the gauge transformations 
that gets the maximal score.

This would mean in some sense a slight 
simplification of our assumption.

\subsection{Could we even predict 
the manifold?}

Very speculatively - and with the success
of predicting the dimension in mind - 
we could seek to argue that the group
of gauge transformations $ {\hbox{{
\huge\bf  
$\times$}}}_x G(x) =
G\times G \times \cdots \times G$
in some way could be claimed to represent
a somewhat larger group than just this
$ {\hbox{{\huge\bf  
$\times$}}}_x G(x) =
G\times G \times \cdots \times G$
in as far as we even on the same representation space of a direct sum of the 
representations $F$ for the different 
points in space time could claim to 
represent {\em also a diffeomorphism 
group}. Since this diffeomorphism 
group shuffles around the direct sum 
of the $F$-type representations we could 
claim, that we managed to represent a 
group which is really the combination 
of the diffeomorphism group and the 
group of gauge transformations on just 
the same space of linear representations 
as the group of gauge transformations 
alone gets represented on as its ``record 
(in our game) representation''. 
Intuitively this means that we have got 
an even bigger group relative to the 
representation than if we just represent 
the Standard Model group $S(U(2)\times 
U(3))$ on its $F$'s. Thus including 
such a diffeomorphism extension sounds 
like providing a superwinner superseding
the formal winner itself the $S(U(2)\times
U(3))$ (or its cross products with 
itself). So there is the hope that 
formulating the details appropriately 
we could arrange to get our true 
prediction become {\em the group 
of gauge transformations with the gauge 
group $S(U(2)\times U(3))$ extended with 
a diffeomorphism group}. If indeed in 
addition the dimension $d=4$ for space 
time favoured by our game, because of its
gauge group for general relativity, and
thus hopefully the group of 
diffeomorphisms for just a four 
dimensional manifold  would get 
exceptionally high score, it becomes 
very reasonable to expect that 
our game could predict just the right 
dimension of the manifold, on which 
the cross product of the standard model
group with itself gets extended by 
the diffeomorphism symmetry. 

This means that we are very close to 
have an argument that the most favoured
symmetry group would precisely be the 
group of Standard Model gauge 
transformations extended by just a four 
dimensional diffeomorphism symmetry.

But if so, it would mean, that we had found 
a principle, a game, favouring precisely 
the group of gauge transformations found 
empirically.

Well, it must here be admitted a little 
caveat: The groups we considered to 
derive the dimension were the group
of Lorentz transformation or Poincare 
transformations, {\em and not the full
group of linear d-dimensional maps as 
would locally correspond to the 
diffeomorphism symmetry}. Thus one should 
presumably rather hope for our scheme
to lead not to the full diffeomorphism 
symmetry as part of the winning symmetry 
group, but rather only that part of the 
diffeomorphism group, which does not shift
the metric tensor $g_{\mu\nu}$. It would 
namely rather be this subgroup of the 
diffeomorphism group that would locally 
be like the $Spin(4)$ or SO(4) as we 
discussed in the dimension fitting.

But somehow this is presumably also
rather what we should hope for to
have a successful theory of ours.

``Going for'' the Standard Model as 
were our starting point means that we 
really concentrated on only looking for 
the long wave length or practically 
accessible part of whatever the true 
theory for physics might be. This long 
wave length practical section should
presumably be defined as what we can 
learn from few particle collisions
with energies only up to about a few 
TeV. But in such few particle practical
experiments we should not discover
gravity and general relativity. We 
should only ``see'' the flat Minkowski 
space time and the Standard Model.
But that should then mean that we should 
not truly ``see'' diffeomorphism group,
but only some rudiments associated 
with the metric tensor leaving part
of this group.

The ideal picture which we should 
hope to become the prediction in this 
low energy section philosophy should 
rather be that the geometrical symmetries
are only the flat Poincare group combined 
with the full gauge group for the Standard
Model.
\section{Conclusion}
\label{conclusion}  
The main point of the present article 
is the suggestion that in a way - that 
may have to be made a bit precise in the 
future/coming further work  - a principle 
of ``small representations'' should be 
sufficient to imply a significant part of 
the details of the Standard Model. The 
real recently most important progress 
in the work with Don Bennett \cite{seeking}
is that it seems that even for the 
selection of the gauge group itself
this selection of ``small representations''
is so important {\em that the very group
is selected so as to in the appropriate 
way of counting have the smallest faithful
representations.} That is to say the 
Standard Model gauge group should  have
been selected to be the model of Nature 
precisely, because it could cope with 
smaller representations, measured in 
our slightly specific way, than any other
proposal for the gauge group (except 
for cross products of the Standard 
Model group with itself a number of 
times). This so successful specific way 
of measuring the ``smallness'' of the 
representations takes its outset from the 
(inverted) Dynkin index in the case 
of simple Lie groups: $C_A/C_F$. This is 
then averaged actually in the way that 
the logarithms of it is averaged weighted
with the dimensions of the various simple
groups in the cross product (and then we
may of course reexponentiate if we want)
and extended to the most natural analogue 
for the Abelian Lie-algebra parts, 
essentially replacing the $C_A/C_F$ 
by $e_A^2/e_F^2$ meaning the charge 
square ratio for two representations 
analogous to the adjoint and the $F$
ones.

The philosophy that taking outset in 
$C_A/C_F$ 
with $F$, as we did, being chosen so as to 
maximize this ratio $C_A/C_F$ can be 
considered assuming a principle of 
``small representations'' is obvious.
If we consider the adjoint representation 
quadratic Casimir $C_A$ for the simple group
under investigation as just a 
normalization - to have something to 
compare quadratic Casimirs of other 
representations to - maximizing our 
starting quantity $C_A/C_F$ means really 
selecting a (simple) group  according 
to how small faithful representations 
$F$ one can find for it. So it is really 
selecting the group with the smallest 
representations. Here of course then 
the concept of the size of the 
representation has been identified with 
the size of the quadratic Casimir, but 
that is at first a very natural 
identification and secondly, that were the 
one with which we had the success. 
It is also the quadratic Casimir, which 
is connected with natural metric on the
space of unitary matrices in the 
representations. In fact our outset quantity
$C_A/C_F$ becomes the square of the 
ratio of the distance the unitary 
representation matrix moves for an 
infinitesimal motion of the group element
in the adjoint and in the representation
 $F$, wherein by choice of $F$ this 
distance is minimal. So our 
``goal quantity'' which is the appropriate
average of the ratio $C_A/C_F$ and its 
extension to the Abelian parts becomes 
(essentially) the square of the volume 
of the volume of the representation space
- in representations of the $F$'s - and 
the corresponding representation space
using the adjoint representation or 
an analogue of adjoint space 
representation, if Abelian parts are 
present. But the crux of the matter is 
{\em a surprisingly large amount of details
of the Standard Model including its 
Gauge group is determined from a 
requirement of essentially minimizing 
the quadratic Casimirs of the 
 representations}:
\begin{itemize}
\item First the {\bf gauge group} - and 
here we stress {\em group} - $S(U(2)\times
U(3))$ of the Standard Model is {\em 
selected} 
by for our ``goal quantity'' (\ref{goal})
obtaining the highest score 2.95782451 
which is rather tinnily,0.0067,above the 
next (silver medal) 
(not being just a trivial cross product 
including the Standard Model itself), 
namely $U(2)$ (= standard model missing 
the
strong interactions QCD)
2.95115179.

\item {\bf The dimension 4 for space time 
is also {\em selected}} by the Poincare
group getting the highest score for 
approximately the same ``goal quantity'',
which we used for the gauge group.
It must be admitted though that we did 
not treat the Poincare group exactly 
- because it does not have the nice 
finite dimensional representations we 
would like to keep to have as strong 
similarity with the gauge group as 
possible - but instead made the trick of 
making some crude corrections starting 
from the Lorentz group. When using the 
Lorentz group dimension d=3 and d=4 stand
equal. When we correct in reasonably 
``fair'' ways the dimension d=4 (the 
experimental one for practical purposes
in our notation that include the time)
wins by having the highest corrected 
``goal quantity'' for the Lorentz group,
corrected to simulate the Poincare group.
In this sense our principle, which is at 
the end a principle of small 
representations, point to the 
experimentally observed number of 
dimensions d=4. 

\item {\bf The representation of the Higgs 
field} is when we use our ``goal quantity''
inspired way of defining in a very precise 
way numerically the smallest of the 
possible various irreducible 
representations to be the inverse of the 
this ``goal quantity'' for the factor group
$G/K$ = $S(U(2)\times U(3))/K$, for which 
the thought upon representation $R$ is 
faithful. By this we just mean that we 
define $K$ as the (invariant) subgroup, 
the elements of which are represented 
just by the unit matrix in the 
representation $R$. This we then in 
principle go through for all irreducible 
representations 
$R$ for the Standard Model and ask for 
each possible $R$:  what is the ``goal 
quantity'' for the corresponding 
$S(U(2)\times U(3))/K$ (here $K$ depends
on $R$ of course) group. For $R = ({\bf 2},
{\bf 1}, y/2 =1/2)$ this factor group
$S(U(2)\times U(3))/K$ turns out to be 
just $U(2)$ and score ``goal quantity''
for the representation $R = ({\bf 2},
{\bf 1}, y/2 =1/2)$ is just that of the 
group $U(2)$ because it happens that
the $F$ for the $SU(2)$ inside $U(2)$
is just the ${\bf 2}$. Thus the quantity 
to determine to decide on the 
representation  $R = ({\bf 2},
{\bf 1}, y/2 =1/2)$ becomes exactly 
the $\hbox{``goal quantity''}_{U(2)}$
which we knew already were unbeatable 
(except if there should have been 
an irreducible representation 
faithful for  the whole Standard Model 
group, but there is not). Thus assuming 
that the representation is smallest 
meaning, since ``size'' 
= 1/``goal quantity'' for representations
using our scheme, for predict
representations the Higgs which is scalar
and has no anomaly problems should be
that representation that won  $R = ({\bf 2},{\bf 1}, y/2 =1/2)$, and that is 
precisely the representation of the Higgs!
\item The Fermion representations 
all for mass protected Fermions 
(meaning that gauge symmetry would 
have to be broken, spontaneously by 
a Higgs presumably) in order for the 
Fermions to obtain nonzero masses.
This makes them easily make anomalies 
in the gauge symmetries 
(charge conservations). In order that no 
anomalies really occur relations
between the number of species of Fermions
in various representations get severely
restricted. Together with some requirement
of  ``small representations'' it looks
rather suggestive, that the Standard Model
system of particles in a family comes out
just intuitively. In our article 
\cite{CostaRica} we did an attempt
to make the requirement of small 
representations precise in a quite 
different way than in the present 
article- but it were an attempt to 
assume small representations in 
some way at least -and we mainly worked
with the Cartan algebra only. But the 
result was, that the Standard Model
representations came out/were postdicted
for the Cartan algebra at least.   

\item At the end we sought to change 
the point of view as to what group 
should be the one, that shall win the 
game of getting the largest 
``goal quantity'' from being {\em the 
gauge group to be the group of all the 
gauge transformations}. Since it happens
that we had balanced our ``goal 
quantity''so much in order to avoid 
making the dimension of the group of much 
influence the value of 
this ``goal quantity'' had turned out to 
be exactly the same for a group and its 
cross product with itself, ever so many 
times. Since now the group of all
gauge transformations is basically an 
infinite cross product of what we 
usually call the gauge group, it means 
that w.r.t. our competition selecting 
the gauge group or the group of gauge 
transformations makes no difference.
So if we e.g. should think that the 
group of all the gauge transformations
is a more fundamental and well defined 
concept, we are free to choose our scheme
to select that group of gauge 
transformations rather than the gauge 
group. 

But if we are very speculatively 
optimistic we might find some argument
that many cross product factors would 
occur and hope in the long run to get 
a kind of understanding of the gauge 
symmetry on a whole manifold to 
optimistically come out of our game.

Perhaps extension of this point of 
view to the Lorentz (or crudely 
Poincare) group as gauge symmetry should
in later work give a better way of 
arguing for the dimension of space time 
d=4, at the same time getting close 
to general relativity. 
    \end{itemize}   
 
This series of ideas for points resulting 
from some principle or another, but 
presumably best by using our ``goal 
quantity'' (\ref{goal}), shows that 
such a type of principle is close to 
deriving a lot of the structure of the
Standard Model: The gauge group, in the
``group'' included some quantization rule
(\ref{restriction}), the space time 
dimension, the Higgs representation, the 
fermion representations, and more 
doubtfully some argument that we have 
gauge symmetry at all.

In conclusion I think that this 
kind of principle   - a precise making 
of a principle of ``small 
representations'' - could have a very 
good chance to explain a lot of the 
structure of the Standard Model and 
thereby of the physics structure,
we see today!

\subsection{Outlook and speculation 
on
finestructure constants}

If we take the above results of having 
success with ``goal quantity'' related
to the representations $F$
being in fact the representations
of the Standard Model group $(y/2=1/2, 
\underline{2}, \underline{1})$ and
$(y/2 = -1/3, \underline{1}, 
\underline{3})$ to mean that these two
representations represent the dominant 
fields (for the gauge field on a lattice 
say), then it {\em happens} that we 
got an ``important representation''
being the direct sum of these two 
representations. This sum corresponds 
to the $\underline{5}$ of the SU(5) in 
grand unification~\cite{GUT}. If we also took it,
that the involvement of the natural 
measure on the representation space of 
unitary matrices in the definition 
of our successful goal quantity to mean, 
that we should use this distance 
measure on the representations to 
suggest the strength of the gauge 
couplings, we would end up with a 
simulated SU(5)-unification prediction!

We hope that our scheme might suggest 
an {\em approximate} SU(5)-relation 
between the couplings only,  because 
we presumably even would if this should 
work at all for our kind of thinking 
rather at some fundammental/Planck scale
than at an adjustable scale like in conventional Grand Unification. (We hope to 
return to our hopes of obtaining 
approximate $SU(5)$ coupling relations 
at the Planck scale in later works in 
which we should then take into account
that there are also secondary 
representations in the series of 
our smallness and that how much they 
shuld contribute might be something we at
least at first could start fitting and 
playing with).  

\section*{Acknowledgement}
I want to thank the Niels Bohr Institute 
for allowing me to stay as emeritus 
and for support of the travel to Bled,
where this proceedings contribution were 
presented and the participants at the 
workshop there and in addition Svend 
Erik Rugh for helpful discussions.
(Hope Svend might participate in next 
work on these ideas).


\begin{thebibliography}{99}
\bibitem{dimension}
  H.~B.~Nielsen,
  ``Dimension Four Wins the Same Game as the Standard Model Group,''
  arXiv:1304.6051 [hep-ph].


\bibitem{seeking} Don Bennett and H. B.
Nielsen, ``Seeking...'', Contribution to 
the workshop ``Beyond the Standard 
Models'', Bled 2011. 
\bibitem{ORaifeartaigh} O'Raifeartaigh, 
Group Structure 
of Gauge theories,University Press 
Cambridge  (1986))



\bibitem{Tegmark} Max Tegmark, ``On the 
dimensionality of space time'', Class.
 Quantum Grav. 14, 1.69 - 1.75 (1997).  

\bibitem{Ehrenfest} Ehrenfest,P., 1917 
Proc. Amsterdam Acad. {\bf 20} 200  

\bibitem{Kane}Gordon Kane : http://particle-theory.physics.lsa.umich.edu/kane/modern.html

\bibitem{Norma} Norma Mankoc et al. 
see many contributions to the present
and previuos Bled Proceedings.
\bibitem{Brene}
  H.~B.~Nielsen and N.~Brene,
  ``Spontaneous Emergence Of Gauge 
Symmetry,''
  IN *KRAKOW 1987, PROCEEDINGS, SKYRMIONS 
AND ANOMALIES*, 493-498 AND COPENHAGEN 
UNIV. - NBI-HE-87-28 (87,REC.JUN.) 6p
  H.~B.~Nielsen and N.~Brene,
  ``Skewness Of The Standard Model:
Possible Implications,''
Physicalia Magazine, The
Gardener of Eden, 12 (1990) 157;
  NBI-HE-89-38;
  H.~B.~Nielsen and N.~Brene,
  ``What Is Special About The Group Of The Standard Model?,''
  Phys.\ Lett.\ B\ {\bf 223} (1989) 399.



\bibitem{Rugh}	H.B. Nielsen, S.E. Rugh
and C. Surlykke, Seeking
Inspiration from the Standard Model
in Order to Go Beyond It, Proc. of
Conference held on Korfu (1992)

\bibitem{Casimir} Oliver, David (2004). 
``The shaggy steed of physics: 
mathematical beauty in the physical 
world.'' 
Springer. p. 81. ISBN 978-0-387-40307-6.
    Humphreys, James E. (1978). 
``Introduction to Lie Algebras and 
Representation Theory.'' Graduate Texts 
in Mathematics 9 (Second printing, 
revised ed.). New York: Springer-Verlag. 
ISBN 0-387-90053-5.
    Jacobson, Nathan (1979). ``Lie 
algebras.'' Dover Publications. 
pp. 243–249. ISBN 0-486-63832-4.





\bibitem{Dynkinindex}
Philippe Di Francesco, Pierre Mathieu, David Sénéchal, Conformal Field
Theory, 1997 Springer-Verlag New York, ISBN 0-387-94785-X


\bibitem{parity}
C.D. Froggatt and H.B. Nielsen``Why 
do we have parity violation?''
arXiv:hep-ph/9906466v1, 23 Jun 1999.

\bibitem{RD}
RANDOM DYNAMICS:
  H.~B.~Nielsen,
  ``Dual Strings,'', ``Fundamentals of quark models'', In:
Proc. of the Seventeenth Scott. Univ. Summer School in Physics,
St. Andrews, august 1976. I.M. Barbour and A.T. Davies(eds.),
Univ. of Glasgow , 465-547 (publ. by the Scott.Univ. Summer School in Physics,
1977)
 (CITATION = NBI-HE-74-15;)


\bibitem{CostaRica} C. D. Froggatt, H.B. Nielsen, Y. Takanishi
 ``Neutrino Oscillations in Extended Anti-GUT Model''
Talk given at the Second Tropical Workshop on 
Particle Physics and Cosmology, San Juan , Puerto Rico May 2000.
arXiv: hep-ph/0011168v1

\bibitem{RD2}
H.B. Nielsen, Har vi brug for fundamentale naturlove(in Danish)
(meaning:``Do we need laws of Nature?'')
{\it Gamma 36 page 3-16, 1978}(1. part) and {\it Gamma 37 page 35-46, 1978}
(2. part)


H.B. Nielsen and C. D. Froggatt,
``Statistical Analysis of quark and lepton masses'',
Nucl. Phys. {\bf B164}(1979) 114 - 140.

\bibitem{Foerster}D. F{\o}rster, H.B. Nielsen, and M. Ninomiya,
``Dynamical stability of local gauge symmetry. Creation
of light from chaos.''
Phys. Lett. {\bf B94}(1980) 135 -140


\bibitem{RDrev}
H.B. Nielsen,
Lecture notes in Physics {\bf 181}, ``Gauge Theories of the Eighties''
In: Proc. of the Arctic School of Physics 1982, Akaeslompolo, Finland,
Aug. 1982. R. Raitio and J. Lindfors(eds.). Springer, Berlin, 1983,p.
288-354.

H.B. Nielsen, D.L. Bennett and N. Brene:
``The random dynamics project from fundamental to human physics''.
{\it In:} Recent developments in quantum field theory.
J. Ambjoern, B.J. Durhuus and J.L. Petersen(eds.),
Elsvier Sci.Publ. B.V., 1985, pp. 263-351



\bibitem{Astri} See the ``home page of Random Dynamics'':

http://www.nbi.dk/~kleppe/random/qa/qa.html 




\bibitem{GUT}
     Ross, G. (1984). Grand Unified 
Theories. Westview Press. 
ISBN 978-0-8053-6968-7.

     Georgi, H.; Glashow, S.L. (1974). 
"Unity of All Elementary Particle Forces".
 Physical Review Letters 32: 
438 441. 
Bibcode:1974PhRvL..32..438G. 
doi:10.1103/PhysRevLett.32.438.

     Pati, J.; Salam, A. (1974). 
"Lepton Number as the Fourth Color". 
Physical Review D 10: 275  289.
 Bibcode:1974PhRvD..10..275P. 
doi:10.1103/PhysRevD.10.275.

     Buras, A.J.; Ellis, J.; Gaillard, 
M.K.; Nanopoulos, D.V. (1978). 
"Aspects of the grand unification of 
strong, weak and electromagnetic 
interactions". Nuclear Physics B 135 (1): 
66G 92. 
Bibcode:1978NuPhB.135...66B. 
doi:10.1016/0550-3213(78)90214-6. 
Retrieved 2011-03-21.

     Nanopoulos, D.V. (1979). 
"Protons Are Not Forever". Orbis 
Scientiae 1: 91. 
Harvard Preprint HUTP-78/A062.

     Ellis, J. (2002). 
"Physics gets physical". 
Nature 415 (6875): 957. 
Bibcode:2002Natur.415..957E. 
doi:10.1038/415957b.

     Ross, G. (1984). 
Grand Unified Theories. Westview Press. 
ISBN 978-0-8053-6968-7.

     Hawking, S.W. (1996). A Brief 
History of Time: The Updated and 
Expanded Edition. (2nd ed.). 
Bantam Books. p. XXX. ISBN 0-553-38016-8.








\bibitem{RDDon1}
H.B. Nielsen and D. L. Bennett,
``The Gauge Glass: A short review'',
Elaborated version of talk at the Conf.
on Disordered Systems, Copenhagen, September 1984. Nordita preprint 85/23.

\bibitem{Baez} see e.g. John Baez: 
http://math.ucr.edu/home/baez/renormalizability.html, November 2006 

\bibitem{Candelas1}P. Candelas, Gary T. Horowitch, Andrew Strominger, 
and Edward Witten, ``Vacuum configurations for superstrings'',NFS-ITP-84-170.

\bibitem{Candelas2}Candelas...Witten, Nuclear Physics B 258: 
46  \% G \% \@ 74,
Bibcode:1985NuPhB.258...46C, doi:10.1016/0550-3213(85)90602-9


\bibitem{suext} See e.g. John M. Pierre, 
``Superstrings Extra dimensions'' 
http://www.sukidog.com/jpierre/strings/extradim.htm


\bibitem{Damgaard} P.H. Damgaard, N. Kawamoto, K. Shigemoto: 
Phys. Rev. Lett.53, 2211 (1984)





\bibitem{MacFarlaine}
A J Macfarlane and Hendryk Pfeiffer,
J. Phys. A: Math. Gen. 36 (2003) 2305
� \ 200 \ 2232317
PII: S0305-4470(03)56335-1
Representations of the exceptional and
other
Lie algebras with integral eigenvalues
of the Casimir operator

\bibitem{Rittenberg}
T. van Ritbergen, A. N.  Shellekens,
J. A. M. Vermaseren,
UM-TH-98-01 NIKHEF-98-004 Group theory
factors for Feynman ...
www.nikhef.nl/~form/maindir/oldversions/.
../packages/.../color.ps

\bibitem{gaugeglas}

Physics Letters B
Volume 208, Issue 2, 14 July 1988, Pages 275-280,
http://arxiv.org/abs/hep-ph/9311321

http://arxiv.org/abs/hep-ph/9607341

Physics Letters B
Volume 178, Issues 2-3, 2 October 1986, Pages 179-186

H.B. Nielsen and N. Brene, Gauge Glass, Proc. of the XVIII International
Symposium on the Theory of Elementary Particles, Ahrenshoop, 1985
(Institut fur
Hochenergiphysik, Akad. der Wissenschaften der DDR, Berlin-Zeuthen, 1985);

\bibitem{Lehto}
Lehto, M., Nielsen H. B., and Ninomiya, M. (1986). Pregeometric quantum lattice: A general discussion. Nuclear Physics B, 272, 213-227.
Lehto, M., Nielsen H. B., and Ninomiya, M. (1986). Diffeomorphism symmetry in simplicial quantum gravity. Nuclear Physics B, 272, 228-252.
Lehto, M., Nielsen H. B., and  Ninomiya, M. (1989). Time translational symmetry. Physics Letters B, 219, 87-91. 





\bibitem{Lehto2}
M. Lehto, H. B. Nielsen, and Masao Ninomiya
``A correlation decay theorem at high temperature''
 Comm. Math. Phys. Volume 93, Number 4 (1984), 483-493. 




\end{thebibliography}
\end{document}